
\input harvmac.tex
\def\pr{\prime}
\def\ol{\overline}
\def\f#1#2{#1\over #2}

\def\ra{\rightarrow}

\def\alp{\alpha}
\def\eps{\epsilon}
\def\gam{\gamma}

\def\lam{\lambda}
\def\Lam{\Lambda}

\def\s{\sigma}
\def\th{\theta}

\def\lp{\ell^\prime}
\def\lpo{\ell^\prime_1}
\def\lpt{\ell^\prime_2}
\def\lpp{\ell^\prime_1+\ell^\prime_2}
\def\lpm{\ell^\prime_1-\ell^\prime_2}
\def\lt{\tilde\ell}
\def\ltp{\tilde\ell^\prime}
\def\lh{\hat\ell}
\def\lhp{\hat\ell^\prime}

\def\sunc{SU(n)/(SU(n-1)\times U(1))}
\def\suthc{SU(3)/(SU(2)\times U(1))}
\def\suoo{SU(1,1)}

\Title{\vbox{\baselineskip15pt\hbox{USC-92/012}}}
{\vbox{\centerline{Modular Invariants of $N=2$ Supersymmetric}
                            \vskip2pt\centerline{$SU(1,1)$ Models}}}

\vskip .30in
\centerline {Katri Huitu\footnote {$^*$}{Supported in part by a grant from the
Emil Aaltonen Foundation}\footnote {$^{\dagger}$}{Address after June 1:
Univ. of Helsinki, Research Institute for High Energy Physics,
Siltavuorenpenger 20 C, 00170 Helsinki, Finland}}

\bigskip\centerline {Department of Physics }
\centerline {University of Southern California }
\centerline {University Park }
\centerline {Los Angeles, CA 90089-0484}

\vskip .3in

We study the modular invariance of $N=2$ superconformal $SU(1,1)$ models.
By decomposing the characters of Kazama-Suzuki model $SU(3)/(SU(2)\times U(1))$
into an infinite sum of the characters of $(SU(1,1)/U(1))\times U(1)$ we
construct
modular invariant partition functions
of $(SU(1,1)/U(1))\times U(1)$.

\Date{5/92}

\newsec{Introduction}

$N=2$ superconformal theories with $c>3$ are  phenomenologically
attractive  as  potentially realistic internal sectors of string
compactifications.
Two types of models of this kind are known:  the Kazama-Suzuki models
based on coset construction of affine Lie algebras
\ref\KS{Y. Kazama, H. Suzuki, Nucl. Phys. B321 (1989) 232.}\
and non-compact models
\ref\DPL{L. Dixon, M. Peskin, J. Lykken, Nucl. Phys. B 325 (1989) 329.},
\ref\IB{I. Bars, Nucl. Phys. B334 (1990) 125.},
\ref\BN{I. Bars, D. Nemeschansky, Nucl. Phys. B348 (1991) 89.}.
In \DPL\
it was shown that all  unitary $N=2$ models with $c>3$ can be written as
\eqn\lcoset{{\f {SU(1,1)}{U(1)}}\times U(1) \, .}

A stringent restriction for conformal theories comes from
 modular invariance.
The modular properties of characters of
 Kazama-Suzuki models without fixed points of  field identifications are well
known
\ref\SY{A. N. Schellekens, S. Yankielowicz, Int. J. Mod. Phys. A5 (1990)
2903.}.
On the other hand
 the modular invariance remains largely an open question  for non-compact
models.
Some attempts to answer these questions
for strings on $SU(1,1)$ manifold  have been investigated in
\ref\HHRS{M. Henningson, S. Hwang, P. Roberts and B. Sundborg, Phys. Lett.
B 267 (1991) 350.}.
It was suggested that the characters of the discrete series $D^\pm$ contain
new sectors, which correspond to  isomorphic algebras with integer twists.
It was argued that then the left-right symmetric product of the
$D^+\oplus D^-$ characters
is a modular invariant.

Another way of finding  modular invariant partition functions for the
$SU(1,1)$ models
is to express a known modular invariant partition function in terms of
non-compact characters.
For instance one could  use this approach to determine the field content of
the
parafermionic coset model
$SU(1,1)/U(1)$ at level $9/4$ or central charge
$c=26$ \BN.
This model is of great interest, since it was identified by Witten
\ref\EW{E. Witten, Phys. Rev. D44 (1991) 314.}\
as a two-dimensional
black hole\footnote*{The string theories in curved space-time were also
discussed in
\BN, and it was
proposed that $SO(d-1,2)_{-k}/SO(d-1,1)_{-k}$ describe strings propagating in
$(d-1)+1$ dimensions.}.
The coset structure \lcoset\ indicates that the $N=2$ partition function of
$SU(1,1)$
can be expressed in terms of the parafermionic partition function
and the bosonic partition function.

As an example of these ideas we study   the
$N=2$ superconformal $SU(1,1)$ models.
In particular, we show  that the characters of the
Kazama-Suzuki model $\suthc $ can be expressed as an infinite sum of
 characters of $SU(1,1)$ model.
Using the relationship between these two theories it is clear that a modular
invariant
combination of the $SU(1,1)$ characters is
given by the diagonal $\suthc $ partition function.

\newsec{$N=2$ Characters of $SU(1,1)$}

It is well known that  minimal $N=2$ models with $c<3$ are closely related to
the
$SU(2)_k$ Kac-Moody algebra
\ref\ZFt{A. B. Zamolodchikov, V. A. Fateev, Sov. Phys. JETP 63 (1986) 913.},
\ref\zQ{Z. Qiu, Phys. Lett. 198B (1987) 497.}.
Both theories can be constructed by combining a free scalar field with
parafermions.
In \DPL\ it was shown that a
similar relationship exists between  $N=2$ models with $c>3$ and $SU(1,1)$
models.
In this case we  combine the generalized parafermions $SU(1,1)/U(1)$
\ref\jL{J. Lykken, Nucl. Phys. B313 (1989) 473.}\
with a free scalar field.
The currents satisfying the operator product expansion of the $SU(1,1)$
Kac-Moody algebra are given by
\eqn\suoocu{\eqalign{
J^+(z) & = \sqrt{k}\psi (z)e^{\sqrt{\f 2k}\phi } ,\cr
J^-(z) & = \sqrt{k}\psi^\dagger (z)e^{-\sqrt{\f 2k}\phi }, \cr
J^3(z) & = -\sqrt{\f k2}\del_z\phi , \cr
}}
where $\psi_1 (z) $ and $\psi_1^\dagger (z)$ are the parafermions and the
field $\phi (z)$
describes the $U(1)$ part.
The supersymmetry generators are obtained by changing the
coefficients of the exponential term in \suoocu,
\eqn\susycu{\eqalign{
G^+(z) & = \sqrt{\f{2c}3}\psi (z)e^{{\sqrt{\f 3c}}\phi } ,\cr
G^-(z) & = \sqrt{\f{2c}3}\psi^\dagger (z)e^{-\sqrt{\f 3c}\phi }, \cr
J(z) & = i\sqrt{\f c3}\del_z\phi . \cr
}}
Using the parafermion algebra \jL\ one can verify that the fields
$G^+,\,G^-,\,J$
of \susycu,
along with the stress tensor
$T(z)=T_\psi -{\f 12} (\del_z\phi )^2$, satisfy the $N=2$ algebra.
The relationship between the central charge $c$ of the Virasoro algebra
and the level $k$  Kac-Moody algebra is  $c=3k/(k-2)>3$.

With this connection between $SU(1,1)$ model and generalized parafermionic
theory,
it is clear that
every field in the theory can be written as a product of a field in the
parafermionic and
a field in the $U(1)$ theory.
The field in the corresponding $N=2$ theory is found by changing the radius of
the boson \DPL.
Thus, the characters of $N=2$ superconformal unitary models with $c>3$ can
be calculated
by first finding the character for the unitary $SU(1,1)/U(1)$ model and then
combining this parafermionic character with a free boson character to form
a character of the unitary $N=2$ representation.
The characters for the parafermionic coset model $SU(1,1)/U(1)$ has been
calculated
in \ref\pG{P. Griffin, Nucl. Phys. B356 (1991) 287;
K. Sfetsos, Phys. Lett. B271 (1991) 301.}.
When one adds back the boson to complete the $N=2$ representation it is
important  to notice
that the $SU(1,1)$ character does not truncate
\jL\
as it does in the compact case of
 $SU(2)_k$, where a $Z_k$ symmetry is present.

Alternatively one can find the characters for representations of $N=2$
superconformal highest weight representations for $c>3$ and view them as
$SU(1,1)$
representations.
The characters from null state structure of unitary $N=2$ representations
have been
derived in \ref\vD{V. K. Dobrev, Phys. Lett. 186 B (1987) 43.},
\ref\yM{Y. Matsuo, Prog. Theor. Phys. 77 (1987) 7930.},
 \ref\eK{E. Kiritsis, Int. J. Mod. Phys. A3 (1988) 1871.}.
We use the results in \ref\BK{I. Bakas, E. Kiritsis, UCB-PTH-91/44. },
where the $N=2$ characters in $NS$ sector were interpreted as non-compact
characters.
The character for a highest weight $N=2$ representation $R$
is defined as
\eqn\chnt{\chi_R(\tau, z ,u=0)={\rm Tr}_R[q^{L_0}w^{ J_0}],}
where $q=e^{2\pi i\tau }$ and $w=e^{2\pi i z }$.
Later we specialize to the characters $\chi_R(\tau ,0,0)$.

The dimension $h_{j,m}$ and the $U(1)$ charge $Q_{j,m}$ of an $N=2$ primary
state are given by
\eqn\dimq{h_{j,m}={\f {-j(j-1) +m^2}{(k-2)}},\quad Q_{j,m}=-{\f {2m}{k-2}}.}
Corresponding to $SU(1,1)$ representations
\ref\bvB{V. Bargmann, Ann. Math. 48 (1947) 568,
Commun. Pure Appl. Math. 14 (1961) 187.}\ there are four possibilities ($NS$
sector):

1. Trivial representation for which $j=0, \, m=0$ and
\eqn\suotch{\chi^{triv} (\tau , z)=
F(q,w){\f {1-q}{(1+q^{1/2}w)(1+q^{1/2}w^{-1})}}\, ,}
where $F(q,w)=\prod_{n=1}^\infty (1-q^n)^{-3}\sum_{n\in {\bf Z}}q^{n^2/2}w^n$.

2. Positive discrete series $D_n^+$, for which $j=n+\eps $, $\eps =0,1/2$ and
$m=j+r$, $r\in {\bf Z}_0^+$ and
\eqn\suodch{\eqalign{&j\ne k/2: \quad \chi^+_{j ,m}(\tau , z)=q^{h_{j,m}}
w^{Q_{j,m}} {\f {F(q,w)}{1+q^{r+1/2}w^{-1}}}\, ,\cr
&j=k/2: \quad \chi^+_{k/2 ,m}(\tau , z)=q^{h_{k/2,m}}
w^{Q_{k/2,m}} {\f {F(q,w)(1-q)}{(1+q^{r+1/2}w^{-1})(1+q^{r+3/2}w^{-1})}}\,
.\cr}}

3. Negative discrete series $D_n^-$, for which $j=n-\eps $, $\eps =0,1/2$ and
$m=-j-r$, $r\in {\bf Z}_0^+$ and
\eqn\suodcn{\eqalign{&j\ne k/2: \quad \chi^-_{j ,m}(\tau , z)=q^{h_{j,m}}
w^{Q_{j,m}} {\f {F(q,w)}{1+q^{r+1/2}w}}\, ,\cr
&j=k/2: \quad \chi^-_{k/2 ,m}(\tau , z)=q^{h_{k/2,m}}
w^{Q_{k/2,m}} {\f {F(q,w)(1-q)}{(1+q^{r+1/2}w)(1+q^{r+3/2}w)}}\, .\cr}}

4. Continuous series, for which $j(j-1)<\eps (\eps -1)$.
In the principal series
$j=1/2+i\rho $, $\rho\in{\bf R}$ and $m=m_0+n$, $n\in{\bf Z},\,|m_0|\leq 1/2 $.
In the supplementary series $j$ is real with $0<|j-1/2|<\eps +1/2$ and
$m=m_0+n$,
$n\in{\bf Z},\,|m_0\pm 1/2|<|j-1/2| $ and
\eqn\suocch{\chi^c_{j,m}(\tau , z)=q^{h_{j,m}}w^{Q_{j,m}}F(q,w) \, .}

We are interested in finding a modular invariant partition function
\eqn\minvpa{Z=\sum_{p,q}N_{p,q}\chi_p(\tau ,0,0)\chi_q(\tau ,0,0)^*.}
We also require that the coefficient of the vacuum contribution in
eq. \minvpa\
is one and the coefficients $N_{p,q}$ are non-negative integers.
Note that in the partition function the characters are specialized to values
$z=u=0$.
In the following we  use these specialized characters.
It is  also useful to define the supersymmetric partition $p(\ell)$ as the
multiplicity of
states that
are generated by operating with the modes of $N=2$
generators on the vacuum
\eqn\spart{F(q,0)={\th_3\over \prod (1-q^n)^3}=\sum p(\ell)q^{\ell }
=1+\sqrt{q}+3q+6q^{3/2}
+11q^2+\dots .}
The simple structure of the characters \suotch -\suocch\ makes it easy
to express $q^n,\, n\in{\bf Z}$ as a sum of the characters.
The relation  is found by first noticing
\eqn\idexp{1=\chi^{triv}(\tau ) +
q-\sum_{t=1/2}^\infty \left( \sum_{s=0}^{2t}(s+1)(-1)^sp(t-s/2)\right)
(q^t -q^{t+1}),\quad  }
where $t\in {\bf Z}/2$ and $s\in {\bf Z}$ and the powers of $q$ on the right
hand
side of eq. \idexp\ are given by
\eqn\qexp{q^{\lp +h_{j,m}}=\chi_{\lp +h_{j,m}}^c (\tau )+\sum_{\l= \lp
+1/2}(-1)^n \left( \prod^n_{i\geq 1}p(\ell_i )\right) \chi_{\ell+h_{j,m}}^c
(\tau )  .}
Here $n$ is defined through $\sum_{i=1}^n \ell_i=\ell-\lp $ and
$\lp,\ell ,\ell_i\in{\bf Z_+}/2$.
The possible values of $j,\, m$, and $h_{j,m}$  are given in
eqs.  \dimq -\suocch.
The $q$ expansion for discrete series gives
\eqn\discexp{ q^{h_{j,m}}= \chi^\pm_{h_{j,m}}-\sum_{t=1/2}^\infty
\left( \sum_{0\leq s\leq t/(r +1/2)} p(t-s(r+1/2)) (-1)^s \right)
q^{t+h_{j,m}},\quad j\ne k/2 ,}
where the powers of $q$ on the right hand side are given either by
eq. \qexp\ or
recursively by eq. \discexp.
Using  eqs. \idexp\ - \discexp\ enables us to express the character of a
 Kazama-Suzuki model
with known modular properties as a sum of non-compact characters.

\newsec{$\suthc $  Characters}

The simplest $N=2$ superconformal model with central charge greater than
three is
the  Kazama-Suzuki model  $\suthc $
with $c=6k/(k+3) $ and  $k>3$.
The character for these models has been
found in  \ref\HN{K. Huitu, D. Nemeschansky, preprint USC-92/011.}.
In $NS$ sector we have
\eqn\char{\eqalign{\chi^{\Lambda} _{\lambda} (\tau ,z) =
\left( {\f {\th_3 (\tau ,z) } {\prod (1-q^n )^3}}\right)^2
\sum_{\sigma \in W} \epsilon (\sigma )
&\sum_{\sigma (t)} q^{{\f 1{k+3}} \left( (\Lambda +t+\rho)^2 -\rho^2
-((\lambda +\hat\rho)^2 -{\hat\rho}^2)\right)}w^Q\cr
&\times \prod_{i=1,2}  {\f 1 {{1+q}^{2(\sigma (\Lambda+\rho+t)-(\lambda
+\hat\rho))\cdot\mu_i}}w^{2(\alp _1-\alp_2)\cdot \mu_i}}. \cr}}
where $\Lam $ ($\lam $) is the highest weight and  $\rho $ ($\hat\rho $) is
half of
the sum of positive roots of $SU(3) $ ($SU(2)\times U(1)$), respectively.
The factor $\epsilon (\sigma) $ is $-1$ for odd elements of the Weyl group
$W$ and $+1$
for even elements.
The subgroup $SU(2)$ has been chosen to lie in the $\alp_1 +\alp_2$ direction,
where $\alp_1$ and $\alp_2$ are the simple roots of $SU(3)$.
In this case the $\mu_i $ are the fundamental weights of $SU(3)$.
The characters with the other boundary conditions for fermions can be obtained
by shifting $z$  by $1/2$, $\tau/2$ or by $\tau /2+1/2$.

To make a connection between $\suthc $ model and $\suoo $ theory,
we need to expand \char\ in powers of $q$.
Consider first  the highest weight $\Lambda=(0,0)$ of $SU(3)$
and the highest weight $\lambda=(0,0)$
of $SU(2)\times U(1)$.
In appendix A we derive the power series expansion of the character
$\chi_{(0,0)}^{(0,0)}(\tau )$,
\eqn\chwqn{\eqalign{\chi^{(0,0)}_{(0,0)}(\tau )&=\sum_{\sigma\in W}
\eps (\sigma )\sum_{quadrants}\eps_q
\{( \sum_{\ell \geq\delta_{\sigma,q}\atop\ell  \in {\bf Z}/2}
\sum_{\lp \geq\xi_{\sigma,q}\atop\lp  \in {\bf Z}/2}  +
 \sum_{\ell \geq\delta^\pr_{\sigma,q}\atop\ell \in  {\bf Z} /2
+ {\f {3(k+3){\bf Z}_+^2}4}}
\sum_{\lp \geq\xi^\pr_{\sigma,q}\atop\lp  \in {\bf Z} /2 +
{\f {(k+3){\bf Z}_+^2}4}})\hat g_1(\ell ,\lp ) \cr
&+
\sum_{\ell\geq\alpha_{\sigma,q}\atop\ell \in {\bf Z}/2}
\sum_{\lp \geq\beta_{\sigma,q}\atop\lp  \in {\bf Z}/2}
\hat g_2(\ell ,\lp )\}\, q ^{\ell+\lp } .\cr}}
As an example, take the $SU(3)$ coset model at level 4 in terms of the
$SU(1,1)$
characters at level 16.
We find $\chi_{(0,0)}^{(0,0)}(\tau )$ in terms of
$SU(1,1)$ characters by using \chwqn\ together with \idexp\ and \qexp.
The explicit multiplicities of the first few terms are
\eqn\suser{\eqalign{
\chi^{\Lambda=(0,0)} _{\lambda=(0,0)}(\tau ,z)&=\chi^{triv}+\chi^c_{2,0}+
\chi^c_{4,0}+
2\chi^c_{11/2,0}+ \chi^c_{6,0}+\chi^c_{7,0}+2\chi^c_{15/2,0}+3\chi^c_{8,0}+
\chi^c_{9,0}\cr
&+2 \chi^c_{19/2,0}+5\chi^c_{10,0}+2\chi^c_{21/2,0}
+\chi^c_{11,0}+6\chi^c_{23/2,0}+4\chi^c_{12,0}+2\chi^c_{25/2,0}+\dots .\cr}}
The subscripts of the continuous characters correspond to $(h,Q)$.
We have checked with Mathematica that also the $J_0$ parts on the left and
right hand
side of eq. \suser\ are
equal.
This is  expected, since if one makes a twist, which preserves the
algebra \ref\bSS{A. Schwimmer, N. Seiberg, Phys. Lett. 184 B (1987) 191.},
\eqn\ltwist{\eqalign{L_n & \ra L_n+\eta J_n +{\f 16}\eta^2 c \delta_{n,0},\cr
J_n & \ra J_n+{\f 13}\eta c \delta _{n,0}, \cr
G^\pm _r& \ra G^\pm _{r\pm \eta}, }}
to the $N=2$ character $\chi(\tau,0,0)$,
and after that a modular transformation $(\tau,z,u)\ra
(-1/\tau,z/\tau,u+(cz^2)/(6\tau))$
\ref\GW{D. Gepner, E. Witten, Nucl. Phys. B 278 (1986) 493.},
one obtains the following
 relation between the twisted and non-twisted characters:
\eqn\ltwist{\chi(-1/\tau,\eta,0)=\chi^\eta \left (\tau,0,0 )|_{S}\right. . }
This implies that also the $U(1)$ parts in eq. \suser\ are equivalent,
when the characters are evaluated at the same point.

It is also possible to deconstruct the characters of other representations
explicitly
similarly than \chwqn.
This deconstruction is given in appendix A.

\newsec{Modular Invariants and Discussion}

We have demonstrated that $\suthc $ characters can be written as a sum of
$SU(1,1)$ characters.
Since modular transformation matrices are unitary, it is well known that
combining the $\suthc $ left and right sectors diagonally leads to a
modular invariant partition function
\eqn\misuoo{Z(\tau )={\f 12}\sum |\chi^\Lam_{\lam ,\nu}(\tau )|^2 ,}
where $\nu $ denotes the different boundary conditions for the fermions.
This means that also the infinite combination of specialized $SU(1,1)$
characters,
given by the decomposition of $\chi^\Lam_\lam $, is
modular invariant.

The infinite number of $SU(1,1)_{16}$ characters in eq. \suser\ forming the
character of
the $k=4$ Kazama-Suzuki model has a natural explanation.
In \DPL\ it was shown that an infinite number of representations
is needed to construct a modular invariant partition function for $N=2$
theory, if $c>3$.
This generalizes Cardy's result \ref\jC{J. Cardy, Nucl. Phys.
B270 [FS16] (1986) 186.}\
for Virasoro minimal models.
In the case of the $\suthc $ model,
when the $q^{1/24}$ factors of $\eta $ functions and vacuum contribution
$q^{-c/24}$
are properly included in the character \char, the maximum growth of the
partition function in the limit
$\tau \ra i\infty $ is given by the vacuum energy $e^{\pi c {\rm Im} \tau/6}$.
On the other hand, after the modular transformation $\tau\ra -1/\tau $
and the transformation to $\tau \ra i\infty $ region, the maximum growth is
$e^{\pi{\rm Im }\tau }$.
This shows that for $c<6$, the Kazama-Suzuki partition function doesn't
necessarily
have an infinite number of terms, when $k=4$, while for the $SU(1,1)$ model
at $k=16$
an infinite
number of representations is needed.

Because of the GSO projection, the partition function for ten dimensional
Neveu-Schwarz-Ramond model is known to vanish by the Riemann identity.
In \ref\skH{S. K. Han, Phys. Rev. D 39 (1989) 2322.}\ it was shown that there
exists a generalized Riemann identity by which the diagonal four dimensional
partition function, based on compactification  on nine copies
of superconformal minimal models at level 1 vanishes.
Along similar lines one should be able to explicitly show that the
four dimensional
partition function for Kazama-Suzuki models $\sunc $
with $c=9$ vanishes.
We hope to return to this issue later.

\vskip0.5in
\noindent
{\bf {Aknowledgement:}}
\vskip0.2in

I would like to thank Dennis Nemeschansky for  suggestions and discussions.
\vskip0.5in

\appendix {A}{}

In this appendix we write the character
as the power series expansion \chwqn.
The character \char\ can be written as
\eqn\chnew{\eqalign{q^{{\f 1p}(\Lam +\rho)^2} &
\sum_{\ell ,\ell^\prime=0}^{\infty}p(\ell )p(\ell^\prime)
q^{\ell+\ell^\prime}\sum_\sigma \eps (\sigma) \left( \prod_{i=1,2}
(\sum_{\ell_i\succ 0 \atop s_i\geq 0 } -
\sum_{\ell_i\prec 0 \atop s_i< 0 } ) \right) (-1)^{s_1+s_2}\cr
& \times q^{p(\ell_1^2 +\ell_2^2 -\ell_1\ell_2)
+ 2 \ol\ell\cdot \sigma(\rho) +p\ol\ell\cdot \ol s + [\sigma(\Lam +\rho )
-\lam -\hat\rho ]\cdot\ol s}.\cr}}
Here $\ell_i \succ 0$ means $\ell_i\geq 0$, if
$(\s(\Lam +\rho )-\lam-\hat\rho )\cdot \mu_i >0$, otherwise $\ell_i<0 $.
If $\ell_i\geq 0 $ in the first term, then $\ell_i<0 $ in the second term,
otherwise
$\ell_i\leq 0$.
For our purpose it is useful to diagonalize the exponent in $\ell_1$ and
$\ell_2$.
To do that we write $\ell_1=\ell^\prime_1+\ell^\prime_2$ and
$\ell_2=\ell^\prime_1-\ell^\prime_2$
and $\lpo\pm\lpt \in Z$.

With this change and changing the summation appropriately
each term in the character \chnew\ can be written in the form:
\eqn\abcdel{\eqalign{
\pm q^{{\f 1{6p}}(3a^2 +b^2)} &\sum_{\ell ,\ell^\prime \atop
\ell,\ell^\prime \in Z/2}p(\ell )
p(\ell^\prime)q^{\ell+\ell^\prime}\cr
&\times \sum_{{\lpo\succ 0 \atop -\lpo\prec \lpt \prec\lpo }\atop s_1,s_2
\succ 0}
(-1)^{s_1+s_2}q^{(p({\lpo }^2+3{\lpt }^2)+a\lpo+b\lpt)
+s_1(p(\lpo +\lpt )+c)+s_2(p(\lpo-\lpt)+d)}.\cr }}
We have divided $\lpo$ and $\lpt$ sums into four quadrants according to
the sign of
$\ell_1$ and $\ell_2$.
In $-\lpo\prec \lpt \prec\lpo$ the '$\prec $' means either $\leq $ or $<$:
$\lpo\geq -\lpt $ if $c>0$ and $\lpt\leq\lpo $ if $d>0$.
Similarly $\lpo\succ 0$ is $\lpo\geq 0$ if both $c>0$ and $d>0$, otherwise
$\lpo >0$.

We consider first \chnew\ with $\Lam =(0,0) $ and $\lam =(0,0)$.
The coefficients $a,\,b,\,c$ and $d$ for different Weyl-group elements are
given in table
for the first quadrant, i.e. when $\ell_1 \succ 0$
and $\ell_2 \succ 0$.
In the first quadrant $s_1\geq 0,s_2\geq 0$.

\vskip10pt
\vbox{\offinterlineskip
\hrule
\halign{
&\vrule#& \strut $\,$\hfil #  \hfil &\vrule #&$\,$ \hfil # \hfil&\vrule #&
$\,$ \hfil # \hfil&\vrule #
& $\,$\hfil # \hfil &\vrule #& $\,$\hfil # \hfil&\vrule #&$\,$ \hfil # \hfil&
\vrule #& $\,$\hfil # \hfil\cr
height2pt&\omit&&\omit&&\omit&&\omit&&\omit&&\omit&&\omit&\cr
&&$\,\,\sigma_0$&&$\,\,\sigma_1$&&$\,\,\sigma_2$&&$\,\,\sigma_{21}$&&
$\,\,\sigma_{12}$&&$\,\,\sigma_{121}$&\cr
height2pt&\omit&&\omit&&\omit&&\omit&&\omit&&\omit&&\omit&\cr
\noalign{\hrule}
height2pt&\omit&&\omit&&\omit&&\omit&&\omit&&\omit&&\omit&\cr
&a&&2&&1&&1&&-1&&-1&&-2&\cr
height2pt&\omit&&\omit&&\omit&&\omit&&\omit&&\omit&&\omit&\cr
\noalign{\hrule}
height2pt&\omit&&\omit&&\omit&&\omit&&\omit&&\omit&&\omit&\cr
&b&&0&&-3&&3&&3&&-3&&0&\cr
height2pt&\omit&&\omit&&\omit&&\omit&&\omit&&\omit&&\omit&\cr
\noalign{\hrule}
height2pt&\omit&&\omit&&\omit&&\omit&&\omit&&\omit&&\omit&\cr
&c&&1/2&&-1/2&&1/2&&-1/2&&-3/2&&-3/2&\cr
height2pt&\omit&&\omit&&\omit&&\omit&&\omit&&\omit&&\omit&\cr
\noalign{\hrule}
height2pt&\omit&&\omit&&\omit&&\omit&&\omit&&\omit&&\omit&\cr
&d&&1/2&&1/2&&-1/2&&-3/2&&-1/2&&-3/2&\cr
height2pt&\omit&&\omit&&\omit&&\omit&&\omit&&\omit&&\omit&\cr}
\hrule}
For the other quadrants (we change the $\ell_i \prec 0$ summations to
$\ell_i \succ 0$):
\eqn\quadrants{\eqalign{&2.\quad\ell_1\prec 0,\,\ell_2\succ 0:
\quad s_1>0,s_2\geq 0;\quad c\ra -c\cr
&3.\quad\ell_1\succ 0,\,\ell_2\prec 0:\quad s_1\geq0,s_2> 0;\quad d\ra -d\cr
&4.\quad\ell_1\prec 0,\,\ell_2\prec 0:\quad s_1>0,s_2> 0;\quad c\ra -c,\,d
\ra -d.\cr}}

We want to write this in a form from which we can read off the multiplicities
of
various $\suoo $ characters.  Let's define
$\ell+s_1[(\lpp)p+c]=\lt $.
 From this it is easy to see that
\eqn\ltboun{s_1\leq {\lt \over (\lpp)p+c}}
and the $\ell$ and $s_1 $ sums can be changed as follows
\eqn\ltsum{\eqalign{\sum_{\ell\succ 0} & \sum_{s_1\succ 0} p(\ell )(-1)^{s_1}
q^{\ell+s_1[(\lpp)p+c]}\cr
&=\sum_{\lt\geq\alpha\atop \lt\in{\bf Z}_+/2}
\sum_{0\prec s_1\leq {\lt\over (\lpp)p+c}} (-1)^{s_1}
p(\lt - s_1[(\lpp)p +c]) \,q^{\lt }\equiv p_1(\lt,\lpo,\lpt) \,q^{\lt} ,\cr}}
where
\eqn\dalp{\alpha = \left\{ {0,\,\,\,\,{\rm if }\,s_1\geq 0 \atop p+c ,\
,{\rm if }\,s_1>0}.\right. }
Similar considerations for $\ell^\prime $ and $s_2$ summations give
\eqn\ltpsum{\eqalign{\sum_{\ell^\prime\geq 0} & \sum_{s_2\succ 0} p(\ell )
(-1)^{s_2}
q^{\ell^\prime+s_2[(\lpm)p+d]}\cr
&=\sum_{\ltp\geq\beta \atop \ltp\in{\bf Z}_+/2} [\sum_{0\prec s_2\prec
{\ltp\over (\lpm)p+d}} (-1)^{s_2}
p(\ltp - s_2[(\lpm)p +d]]\, q^{\ltp }\equiv p_1^\prime(\ltp,\lpo,\lpt)\,
q^{\ltp} ,\cr}}
where
\eqn\dbeta{\beta = \left\{ {0,\,\,\,\,{\rm if }\,s_2\geq 0 \atop p+d ,\,
{\rm if }\,s_2>0}.\right. }
We still want to deal with the summations over $\lpo$ and $\lpt$.

Consider first the case when $\lpo$ and $\lpt $ are integers and
$\lpo \ne 0$.
(Note that they are always both either integers or integers+half.)
The dependence of $\lpt$ is
\eqn\lptdep{\sum_{-\lpo\prec \lpt\prec\lpo} \sum_{\lt \geq\alpha}
p_1(\lt,\lpo,\lpt) p_1^\prime(\ltp,\lpo,\lpt) q^{\lt +3 {\lpt }^2p+b\lpt}.}
Changing summation over the variable $\lh $, where $\lh $ is defined as
\eqn\lhdef{3 {\lpt }^2p+b{\lp}_2 +\lt \equiv \lh \in {\bf Z}/2}
we get \lptdep\ to the following form
\eqn\lhsum{\sum_{\lh\geq \delta \atop \lh \in {\bf Z}/2}
\sum_{\tilde\gam_-\prec\lpt\prec\tilde\gam_+} p_1(\lt,\lpo,\lpt)
 p_1^\prime(\ltp,\lpo,\lpt) \,q^{\lh}
\equiv \sum_{\lh\geq \delta \atop \lh \in {\bf Z}/2} g(\lh,\ltp,\lpo ) \,
q^{\lh} .}
Here  $\tilde\gam_\pm=
\{{\min\atop\max}\}\{\pm\tilde\lpo,-b/6p \pm\sqrt{b^2+12p\lh-12\alpha p}/6p\}$
 and
$\tilde\lpo=\left\{\lpo,\,\,{\rm if }\,\, \lpo\geq \pm\lpt \atop \lpo-1,\,
\,{\rm
otherwise} \right. $ and $\delta =\alpha -b^2/12p $.

Defining $ \ltp +{\lpo }^2 p +a \lpo =\lhp $, we get from the $\lpo $ summation
\eqn\lhpsum{\sum_{\lhp\geq\xi\atop\lhp  \in {\bf Z}/2}
\sum_{\max\{\tilde\zeta,\zeta_-\}\prec\lpo\prec\zeta_+}\, g(\lh,\ltp,\lpo )
 \,q^{\lhp}
\equiv  \sum_{\lh \geq\delta\atop\lh  \in {\bf Z}/2}
\sum_{\lhp \geq\xi\atop\lhp  \in {\bf Z}/2} \hat g_1 (\lh,\lhp) q^{\lhp} .}
Here $\xi=\beta-a^2/4p$, $\zeta_\pm=a/2p\pm\sqrt{a^2+4p\lhp-4p\beta }/2p$ and
$\tilde\zeta =\left\{ {0,\,{\rm if}\, \lpo\geq 0
\atop 1,\,{\rm  otherwise}} \right. $.

 From the above we have that when $\lpo$ and $\lpt $ are integers and
$\lpo \ne0$,
 the contribution to the character is
\eqn\firstint{\sum_{\lh \geq\delta\atop\lh  \in {\bf Z}/2}
\sum_{\lhp \geq\xi\atop\lhp  \in {\bf Z}/2} \hat g_1 (\lh,\lhp) q^{\lh +\lhp} }

If $\lpo$ and $\lpt $ are integers+half we get otherwise similar term, but
$\lh \geq \max (\delta ,{\f {3p}4}+{\f b2})\equiv\delta ^\pr,\,\lh\in
{\bf Z} /2 +
{\f {3pn^2}4}$ and
$\lh \geq \max (\xi ,{\f {p}4} +{\f a2})\equiv\xi ^\pr,\,\lh\in {\bf Z} /2
+ {\f {pn^2}4}$.

If finally $\lpo=0$, it is easy to see that the contribution of this term is
\eqn\zlpo{\sum_{\lh \geq\alpha\atop\lh  \in {\bf Z}/2}
\sum_{\lhp \geq\beta\atop\lhp  \in {\bf Z}/2} \sum_{{0\prec s_1\leq\lh/c \atop
0\prec s_2\leq\lhp/d }} (-1)^{s_1+s_2}p(\lh -s_1 c) \,p(\lhp -s_2 d) \,q^{\lh
 +\lhp }
=\sum_{\lh \geq\alpha\atop\lh  \in {\bf Z}/2}
\sum_{\lhp \geq\beta\atop\lhp  \in {\bf Z}/2} \hat g_2(\lh,\lhp )
q^{\lh +\lhp }.}
Combining the previous results gives for the
character
\eqn\chwq{\eqalign{\sum_{\sigma \in W}\eps (\sigma )&\sum_{quadrants}\eps _q
\{( \sum_{\lh \geq\delta_{\sigma,q}\atop\lh  \in {\bf Z}/2}
\sum_{\lhp \geq\xi_{\sigma,q}\atop\lhp  \in {\bf Z}/2}  +
 \sum_{\lh \geq\delta^\pr_{\sigma,q}\atop\lh  \in {\bf Z} /2 +
 {\f {3p{\bf Z}_+^2}4}}
\sum_{\lhp \geq\xi^\pr_{\sigma,q}\atop\lhp  \in  {\bf Z} /2 +
{\f {p{\bf Z}_+^2}4}})\hat g_1(\lh,\lhp ) + \cr
&\sum_{\lh \geq\alpha_{\sigma,q}\atop\lh  \in {\bf Z}/2}
\sum_{\lhp \geq\beta_{\sigma,q}\atop\lhp  \in {\bf Z}/2}
\hat g_2(\lh,\lhp )\}\, q ^{\lh +\lhp } ,
\cr}}
where the subscript $q$ refers to specific values in the given  quadrant.
$\eps _q=1$ for the first and fourth quadrant, otherwise it is $-1$.
For general $\Lam,\,\lam $ the summation indices $\lh ,\lhp$ contain the
fractional part
 from the coefficients $a,b,c,d$.

\listrefs

\bye